\documentclass[a4paper,12pt]{article}
\usepackage{cite}
\usepackage{amssymb}
\usepackage{amsmath}
\usepackage{epsfig}
\usepackage{graphicx}
\usepackage{slashed}
\usepackage{xcolor,color}
\usepackage{ulem}
\usepackage{subfigure}
\usepackage[toc,page]{appendix}
\usepackage[makeroom]{cancel}
\usepackage{tikz}
\usetikzlibrary{shapes.geometric,positioning,decorations.pathreplacing,decorations.pathmorphing,patterns,decorations.markings}

\makeatletter
\renewcommand{\theequation}{\thesection.\arabic{equation}}
\@addtoreset{equation}{section}
\makeatother
\def\be{\begin{equation}}
\def\ee{\end{equation}}
\def\bea{\begin{eqnarray}}
\def\eea{\end{eqnarray}}
\def\bgn{\begin{align}}
\def\egn{\end{align}}

\def\({\left(}
\def\){\right)}
\def\<{\left<}
\def\>{\right>}

\def\f{\frac}

\def\({\left(}
\def\){\right)}
\def\<{\left<}
\def\>{\right>}
\def\!{\right|}
\def\|{\left|}

\def\[{\left[}
\def\]{\right]}

\def\+{\bar}

\def\ng{{\negthinspace}}

\def\bR{{\bf R}}

\def\l{{{\ell}}}

\def\P{{\cal{P}}}
\def\V{{\cal{V}}}

\usepackage{hyperref}
\hypersetup{colorlinks,%
  citecolor=blue,%
  linkcolor=blue,%
  pdftex}

\begin{document}

\begin{titlepage}
\vskip1cm
\begin{flushright}
% UOSTP {\tt 1512001}
\end{flushright}
\vskip0.25cm
\centerline{
\bf \large 
Janus Deformation of 
de Sitter 
Space and Transitions in Gravitational Algebras %Transition 
} 
\vskip0.8cm \centerline{ \textsc{
 Dongsu Bak,$^{ \negthinspace  a}$  Chanju Kim,$^{ \negthinspace b}$ Sang-Heon Yi,$^{\negthinspace c}$} }
\vspace{0.8cm} 
\centerline{\sl  a) Physics Department \& Natural Science Research Institute}
\centerline{\sl University of Seoul, Seoul 02504 \rm KOREA}
 \vskip0.2cm
 \centerline{\sl b) Department of Physics, Ewha Womans University,
  Seoul 03760 \rm KOREA}
   \vskip0.2cm
 \centerline{\sl c) Center for Quantum Spacetime \&  Physics Department}
  \centerline{\sl Sogang University,  Seoul 04107 \rm KOREA}
\vskip0.4cm

 \centerline{
\tt{(\small dsbak@uos.ac.kr,\,cjkim@ewha.ac.kr,\,shyi@sogang.ac.kr})
} 
  \vspace{1.5cm}
%\centerline{\today}
%\vspace{1.75cm}
\centerline{ABSTRACT} \vspace{0.65cm} 
{
\noindent
We consider a time-dependent $\mathcal{O}(1/G)$ deformation of pure de Sitter (dS)
space in dS gravity coupled to a massless scalar field. It is the dS counterpart
of the AdS Janus deformation and interpolates two asymptotically dS spaces
in the far past and the far future with a single deformation parameter.
The Penrose diagram can be elongated along the time direction indefinitely
as the deformation becomes large. After studying the classical properties
of the geometry such as the area theorem  and the fluctuation by a matter field, 
we explore the algebraic structure of the field operators on the deformed
spacetime. We argue that the algebra is a von Neumann factor of type
II$_\infty$ for small deformations, but there occurs a transition to type 
I$_\infty$ as the deformation increases so that the neck region of the  deformed space becomes 
a Lorentzian cylinder.

}

%\vspace{0.75cm}
%\centerline{(\today)}
\end{titlepage}
%%%%%%%%%%%%%%%%%%%%%%
%\maketitle

%%%%%%%%%%%%%%%%%

%\centerline{\large Januslike Deformation of dS Space and %Gravitational 
%Algebras}

%%%%%%%%%%%%%%%%%%%%%%%%%%%%%%%%%%%%%%%%%%%%%%%%%%%
\section{Introduction
}\label{sec1}
%%%%%%%%%%%%%%%%%%%%%%%%%%%%%%%%%%%%%%%%%%%%%%%%%
De Sitter (dS) space represents the maximally symmetric vacuum solution of the Einstein equations with a positive cosmological constant, %commonly 
associated with the inflationary epoch to the universe~\cite{Starobinsky:1986fx,Goncharov:1987ir,Maldacena:2024uhs}. This space is spatially compact, corresponding to a closed universe,  which prevents especially %in particular  
the construction of the conventional in and out scattering states in  relativistic quantum field theory. 
Consequently, this complicates the interpretation of particle concepts and certain field theory computations.
%Nevertheless, curved space quantum field theory treats the geometry in the $G\rightarrow 0$ limit as a classical background of order ${\cal O}(1/G)$. It is considered capable of describing physics at least locally on weakly curved  dS space 
%Nevertheless, curved space quantum field theory, which treats the geometry in the $G\rightarrow 0$ limit as a classical background of order ${\cal O}(1/G)$, 
%is considered capable of describing physics  at least locally on (weakly curved) dS space\cite{Witten:2001kn}.   
Nevertheless, curved space quantum field theory, which mainly deals with ${\cal O}(1)$ perturbations of graviton and matter fields in a classical background of order ${\cal O}(1/G)$, %which 
may be rigorously defined at least locally~\cite{Witten:2001kn,Bousso:2002fq}.
In the algebraic approach to quantum field theory, one focuses on operator algebras which may provide a more rigorous framework for understanding the entanglement structure of an open region (relative to its complement) and the corresponding (relative) entanglement entropy~\cite{Witten:2018zxz}. 
%The algebraic approach to quantum field theory, focused on its operator algebras, may provide a more rigorous framework for understanding the entanglement structure of an open region (relative to its complement) and the corresponding (relative) entanglement entropy~\cite{Witten:2018zxz}. 
The algebra %type 
of field operators on any open region with a partial Cauchy surface in Minkowski space is believed to be of the %so-called 
von Neumann type ${\rm III}_{1}$ \cite{Haag:1996hvx}. 
%In the context of curved space quantum field theory,  %it is still believed that the algebra %type 
%remains 
In a curved space, 
the algebra %of fields 
is still believed to be
of von Neumann type ${\rm III}_{1}$, even with graviton and various matter fields included
%when massless graviton %field operators 
%as well as various matter fields %operators 
%are included  %in the algebra 
(see~\cite{Witten:2021jzq} for a recent review). Recent %interesting 
developments in the algebraic approach  to semiclassical quantum gravity~\cite{Leutheusser:2021qhd,Leutheusser:2021frk,Witten:2021unn, Strohmaier:2023hhy, Witten:2023qsv, Strohmaier:2023opz, Chandrasekaran:2022cip, Chandrasekaran:2022eqq, Jensen:2023yxy, Kudler-Flam:2023qfl,Kudler-Flam:2023hkl,Faulkner:2024gst,Kudler-Flam:2024psh,Chen:2024rpx,AliAhmad:2023etg} have revealed  
various %some genuine %intriguing 
quantum gravity effects that go %extend 
beyond the conventional %perturbative 
%semiclassical 
framework of the  curved space quantum field theory~\cite{Wald:1995hf,Yngvason:2004uh,Hollands:2014eia,Khavkine:2014mta,Fewster:2019ixc}. In particular, the generalized entropy,  $S_{gen} = A/4G + S_{vN}$ \cite{Bekenstein:1974ax,Bombelli:1986rw,Susskind:1994sm,Wall:2011hj,Faulkner:2013ana,Engelhardt:2014gca},
%, consisting of the area and the von Neumann entropy of quantum fields~
  has been understood
%, in the algebraic context, 
based on a von Neumann type ${\rm II}_{\infty}$  for black holes and a ${\rm II}_{1}$ algebra for pure dS space.

%In  the context of dS quantum gravity,  diffeomorphism invariance introduces  subtleties regarding physical observables even in the $G\rightarrow 0$ limit. Specifically, the isometry acts as a gauge constraint on field operators.  It is argued that the algebra of observables become trivial  when the gauge constraint is imposed  naively on observables on pure de Sitter space. Here, the crucial constraint arises from the non-compact direction of isometries, {\it i.e.} the boost symmetry of the observer's static patch. To obtain  physically sensible results and avoid the triviality of the algebra, the observer's Hamiltonian is introduced and incorporated into the algebra structure. The resulting algebra on the observer's static patch in pure de Sitter space is of the von Neumann type ${\rm II}_{1}$.  Further development in this direction for a generic spacetime  is provided in~\cite{Witten:2023xze}, where  it is argued that  the algebra may realize the background independence of quantum gravity. In addition, the algebra on the timelike envelope of the observer's worldline with a partial Cauchy surface is argued to be a  type ${\rm II}$ von Neumann factor. 
In the context of dS quantum gravity, diffeomorphism invariance introduces subtleties regarding physical observables even in the $G\rightarrow 0$ limit. In particular,  isometries act as gauge constraints on the field operators.  It is argued that the algebra of observables becomes completely trivial if %when 
the gauge constraints are naively imposed on observables on pure dS space. Here a relevant part of %one 
constraints 
comes from a certain non-compact direction  %of isometries, 
%{\it i.e.} 
corresponding to
the boost Killing symmetry 
in the observer's static patch. %In order 
To obtain physically %sensible %
meaningful 
results and to avoid the triviality, %of the algebra, 
the observer's Hamiltonian %further %of the observer 
has to be %is required to be %introduced 
%and 
% further 
incorporated further
into the algebraic structure of the system. The resulting algebra on the %observer's 
static patch 
%of pure dS space 
turns out to be of von Neumann type ${\rm II}_{1}$.  A further progress %development 
in this direction for a generic spacetime is made in~\cite{Witten:2023xze}  (see also \cite{Jensen:2023yxy,Kudler-Flam:2023qfl,Chen:2024rpx,Kudler-Flam:2024psh}), where it is argued  that %the algebra may realize the background independence of quantum gravity. 
the background independence of quantum gravity can be realized in the algebraic context.
In addition, it is further argued that the relevant algebra on the timelike envelope of the observer's worldline with a partial Cauchy surface must be of  von Neumann type ${\rm II}$. % von Neumann factor. 

%In this paper, we consider a classically-deformed space, {\it i.e.} the order ${\cal O}(1/G)$ deformation from pure dS  space by a single scalar field $\phi$.  Especially, we focus on the classical time-dependent solution to the Einstein equation, which starts in far past as dS space and ends  in far future as dS space. Our  time-dependent geometry has a single deformation parameter $\gamma$, which controls the elongation of the geometry along the time direction. We call our solution as Janus deformation of dS space,  since it interpolates two dS spaces with different scalar vevs, and behaves just like the dS counter part of the AdS Janus solutions in~\cite{Bak:2003jk}.  One may note that our solution is completely consistent with the Gao-Wald theorem~\cite{Gao:2000ga}, since our scalar field is an ordinary matter satisfying null energy condition.  After some description of the classical property of our solutions, we explore the algebraic structure of field operators on this background following recent works. We argue that the transition occurs in von Neumann algebra type  from ${\rm II}_{\infty}$ to ${\rm I}_{\infty}$ according to values of  the deformation parameter $\gamma$, or its incarnated parameter $\tau_{0}(\gamma)$. 
In this paper, we consider a classically deformed space, {\it i.e.} an ${\cal O}(1/G)$ deformation of  dS  space by a massless scalar field $\phi$.  In particular, we shall focus on a %classical 
time-dependent solution   %of the Einstein equation, 
which starts in the far past as asymptotically dS space and ends again    as asymptotically dS space in the far future. This %time-dependent 
geometry involves a single deformation parameter $\gamma$ that basically controls the elongation of the geometry along the time direction. We shall call this solution as Janus  deformed dS space,  since it interpolates two asymptotically dS spaces with different scalar vevs %and works 
and behaves as %behaving as %, and corresponds to %just like 
the dS counterpart of the AdS Janus solutions in~\cite{Bak:2003jk,Bak:2007jm}.  One may note that our solution is fully consistent with the Gao-Wald theorem~\cite{Gao:2000ga} as our scalar field 
%is an ordinary matter 
satisfies the null energy condition.  
After studying the classical properties
of the deformed geometry such as the area theorem  and the fluctuation by a matter field, 
%After some descriptions of its classical properties, 
%of our solutions, 
we shall explore the algebraic structure of the field operators  on the deformed
spacetime. %background.
 % following recent work. 
%We argue that the transition occurs in von Neumann algebra types  from ${\rm II}_{\infty}$ to ${\rm I}_{\infty}$ depending on the values of  the deformation parameter $\gamma$, or its incarnated version $\tau_{0}(\gamma)$. 
We shall argue that the algebra is a von Neumann factor of type
II$_\infty$ for small deformations, but there occurs a transition to type 
I$_\infty$ as the deformation increases such that the neck region of deformed space becomes 
a Lorentzian cylinder.

%This paper is organized as follows: In section~\ref{sec2}, we present our solution in two different global coordinates and the interpretation of our solution as a particle motion under a specific potential. After presenting the classification of the solutions, we focus on the specific solution interpolating two de Sitter spaces, Janus deformation of de Sitter space.   In the three-dimensional case,  the explicit formulae of the metric and scalar field are presented.  In section~\ref{sec3}, we describe the observer's patch in our Janus deformed dS space and its horizon area. In the classical context, we identify the entropy of the observer's patch with the horizon area and describes the area law or the increase of the entropy.  Massive scalar field  on the Janus deformed geometry is solved revealing its exponential decaying nature along the time evolution.  In section~\ref{sec4}, the algebra of field operators are considered. The algebra type in the observer's patch for $\frac{\pi}{2} < \tau_{0} <\pi$ is argued to be von Neumann ${\rm II}_{\infty}$ and its transition to type ${\rm I}_{\infty}$ is also illuminated. We conclude in section~\ref{sec5} with some comments. In Appendex~\ref{AppA}, the integral representation of $\tau_{0}$ is expanded to obtain the asymptotic expression of $\tau_{0}$ in terms of $\gamma$. In Appendix~\ref{AppB}, we provide the explicit proof of the area theorem for the horizon area in the observer's patch. 
This paper is organized as follows: 
%In section~\ref{sec2}, we describe our solution in two different global coordinates and give the interpretation  as a particle motion under a specific potential. 
In Section~\ref{sec2}, we consider the $d$-dimensional dS gravity coupled to a massless scalar field. We  classify all possible solutions that preserve SO$(d)$ symmetries.
%After presenting the classification of possible solutions, 
Among them, 
we shall be specialized in %focus on 
%the specific solution that interpolates two dS spaces, 
the Janus deformed geometry and construct the corresponding metric and the scalar field  explicitly.
%In the three-dimensional case,  the explicit formulae of the metric and scalar field shall be presented.  
%In section~\ref{sec3}, we describe the observer's patch in our Janus deformed dS space and its horizon area. In the classical context, we identify the entropy of the observer's patch with the horizon area and describe the area law or the increase of the entropy. 
In Section~\ref{sec3}, we study the global structure of our Janus deformed spacetime. We  find that the Penrose diagram can be elongated along the time direction indefinitely
as the deformation becomes large.  We study various classical properties
of the geometry such as the area theorem, the generalized second law and the fluctuation by a matter field.
%Specialized in the observer's patch at $\theta=0$, we prove the area theorem saying that the area grows monotonically along the observer's future horizon.   
%The massive scalar field  on the Janus deformed geometry is solved verifying its exponential decaying nature along the time evolution.  
%In section~\ref{sec4}, the algebra of field operators are considered. The algebra type in the observer's patch for $\frac{\pi}{2} < \tau_{0} <\pi$ is argued to be von Neumann ${\rm II}_{\infty}$ and its transition to type ${\rm I}_{\infty}$ for $\tau_{0} > \pi$ is also illuminated. 
In Section~\ref{sec4}, we shall examine the algebra of field operators in the deformed spacetime. We shall argue that within the observer's patch for $\frac{\pi}{2} < \tau_{0} < \pi$, this algebra is of von Neumann type ${\rm II}_{\infty}$, and discuss also its transition to type ${\rm I}_{\infty}$ as $\tau_{0}$ becomes larger than $\pi$. Here  $\tau_0(\gamma)$ is the 
range %$\gamma$-dependent 
parameter for the conformal time $\tau \in [-\tau_0, \tau_0]$ (see (\ref{ansatz}) and (\ref{tauzero})). 
%We conclude in 
%Section~\ref{sec5} with some comments. 
Section~\ref{sec5} comprises our conclusions with some comments.
In Appendix~\ref{AppA}, we investigate properties of the parameter $\tau_{0}$ as a function of $\gamma$. %the integral representation of $\tau_{0}$ is expanded %to obtain the asymptotic expression of $\tau_{0}$ in terms of $\gamma$. 
%to the leading order in $\gamma$. 
In Appendix~\ref{AppB}, we give an explicit proof of the area theorem and the generalized second law along the observer's horizon.  

%%%%%%%%%%%%%%%%%%%%%%%%%%%%%%
\section{Deformation of dS Space
}\label{sec2}
%%%%%%%%%%%%%%%%%

In this note, we are interested in a Janus deformation of dS gravity coupled to a massless scalar field. %, which %This 
The system 
is described by the action
\bea
I=%I_{top}+
\frac{1}{16\pi G}\int d^d x \sqrt{-g}\left( R-g^{ab}\partial_a \phi \partial_b \phi -\frac{1}{\ell^2}(d-1)(d-2)\right) \,,
%+ I_{surf}  + I_M(g, \chi) \,,
\label{action}
\eea
where  $\ell$ is the dS radius and %we assume 
the spacetime  dimension $d$ is taken to be greater than or equal to three. % $d \ge 3$.
% $\phi$ is a dilaton field, $\chi$ a matter field and
%
%In this note, 
We shall set $\ell$ to be unity for simplicity of our presentation. 
The equations of motion read
%are obtained from the variation of the metric $g$ and the scalar field $\chi$,
\begin{align}    \label{eom1}
R_{ab} \,\,&=(d-1) g_{ab}+ \nabla_a  \phi \nabla_b \phi\,,\\
 \nabla^2 \phi  &=0\,.  \label{eom2}
\end{align}
For the Janus deformation, we take our ansatz as
\begin{align}   % \label{}
ds^2 = f(\tau)\left(-d\tau^2+ d\Omega^2_{{d-1}}\right)\,,  \ \ \ \phi=\phi(\tau)\,,
%\nonumber \\
% \tanh \frac{t L }{\ell^2} &=\frac{2\sin (\tau-\tau_B)}{(b+b^{-1}) \sin \mu -(b-b^{-1}) \cos (\tau-%\tau_B)}\,,
 \label{ansatz}
\end{align}
where $d\Omega^2_{{d-1}}$ 
 is the line element for the unit $d\ng -\ng 1$ sphere which may be given explicitly by
\be
ds^2_{S^{d-1}}=d\theta^2 + \sin^2 \theta \, %ds^2_{S^{d-2}}
d\Omega^2_{{d-2}}\,.
\label{spherem}
\ee
 In this ansatz, we preserve the SO$(d)$ symmetries %of the $d\ng-\ng 1$ sphere, 
out of  SO$(d,1)$ 
of  pure dS space. 

The scalar equation is solved by
\be
\dot{\phi}=\gamma f^{\frac{2-d}{2}}\,,
\label{sfield}
\ee
where $\gamma$ is our deformation parameter and the dot %$\cdot$ 
represents a derivative with respect to the $\tau$ coordinate\footnote{In the following, we shall restrict our discussion to
$\gamma\ge 0$ since the geometry is invariant under the exchange of $\phi$ and $-\phi$.}\ng. 
The remaining equations of motion can be solved if the conformal factor $f$ satisfies
\be
\dot{f}\dot{f}= 4f^3-4f^2 +\frac{4\gamma^2}{(d-1)(d-2)} f^{4-d} \equiv -V(f)\,.
\label{particle}
\ee
These two equations agree with those of the so-called Janus solutions  \cite{Bak:2003jk,Bak:2007jm,Bak:2007qw} which are spacelike deformations in AdS space. 

As noted before \cite{Bak:2003jk}, the equation (\ref{particle}) may be interpreted as describing the trajectories of a particle whose position is  given by $f \ge 0$. In particular, the 1d particle is moving under the potential $V(f)$ with zero energy. 
Based on this particle picture, one may see that the above equation involves eight classes of solutions.  When $\gamma > \gamma_c$ with $\gamma^2_c=(d-2)\left(\frac{d-2}{d-1}\right)^{d\ng-\ng 2}$\ng\ng,\,\, 
one has $V < 0$ for any $f\in (0,\infty)$ and one has the following two classes. Class I: The particle starts at infinity and ends at $f=0$. In terms of geometry, one begins with an asymptotically dS space in the far past ending up with a big-crunch singularity in the far future. Class II: The particle starts with $f=0$ (a big-bang singularity) and ends  with $f=\infty$ (an asymptotically dS space). When $\gamma=\gamma_c$, the potential has one double zero at
$f=f_0\equiv (d-2)/(d-1)$, for which we have the following four classes.  Class III: The particle starts from $f=f_0$ in the far past  (an asymptotically cylinder spacetime $R_-\times S^{d-1}$ with a constant  kinetic energy density of the scalar field) and goes to infinity (an asymptotically dS space).   Class IV: The particle starts with infinity  (an asymptotically dS space) and approaches $f=f_0$ (an asymptotically cylinder spacetime $R_+\times S^{d-1}$ with a constant kinetic energy density of the scalar field). 
Class V: The particle starts from $f=0$ (a big-bang singularity) and ends with  $f=f_0$ (an asymptotically cylinder spacetime $R_+\times S^{d-1}$ with a constant kinetic energy density of the scalar field). Class VI: The particle starts with $f=f_0$ in the far past  (an asymptotically cylinder spacetime $R_-\times S^{d-1}$ with a constant  kinetic energy density of the scalar field) and goes to $f=0$ (a big-crunch singularity).
When $0 \le \gamma < \gamma_c$, the potential $V(f)$ has two real positive zeros $f_\pm$ with $f_+ > f_- >0$ and the region $f\in (f_-,f_+)$ becomes forbidden classically; In this case, we have the following two classes. Class VII: The particle starts with $f=0$ (a big-bang singularity), reflected back at $f=f_-$ and goes back to $f=0$ (a big-crunch singularity). Class VIII: The particle starts with $f=\infty$ (an asymptotically dS space), reflected back at $f=f_+$ and goes back to $f=\infty$ (another asymptotically dS space).  

Here in this note we shall focus on the last class (with $0 \le \gamma < \gamma_c$ and $f\in [f_+,\infty)$), which is the  case 
%that does not involve any singularities and leads 
leading to an asymptotically dS space in the far past and far future. In this case, (\ref{particle}) can be integrated as
\be
\tau_0-|\tau|=\int^\infty_{f}\frac{dx}{\sqrt{-V(x)}}\,,
\label{timetau}
\ee
where
\be
\tau_0=\int^\infty_{f_+}\frac{dx}{\sqrt{-V(x)}} \,.
\label{tauzero}
\ee
Note that $f(0)=f_+$ and $\tau$ is then   ranged over $[-\tau_0,\tau_0]$.
 %with the choice of integration constant in (\ref{timetau}).
%where one can show that $\tau_0 \ge \pi/2$. 
In Appendix \ref{AppA}, we show that $\tau_0$ begins with $\pi/2$ at $\gamma=0$,
becomes monotonically increasing as a function of $\gamma$ for $\gamma \in [0,\gamma_c)$
and approaches infinity as one takes the limit $\gamma \rightarrow \gamma_c$. In the appendix,  we also show that $\tau_0$ has an expansion
\be
\tau_0=\frac{\pi}{2}  \left(1+ \frac{ (2d\ng -\ng 3) \Gamma(2d\ng -\ng4)}{2^{2d-4}\Gamma(d)\Gamma(d\ng-\ng 1)}\gamma^2 + {\cal O}(\gamma^4)   \right) \,.
\label{tauseries}
\ee 
From $V(f_+)=0$, one may easily verify that  $f(0)$ has an expansion 
\be
f(0)=1- \frac{\gamma^2}{(d-1)(d-2)} + {\cal O}(\gamma^4)    \,.
\label{f0series}
\ee   
Once $f(\tau)$ is given, (\ref{sfield}) can be integrated to give 
\be
\phi(\tau)=\phi_-%\phi(-\tau_0)
+\gamma\int^\tau_{-\tau_0}d\tau'  f^{\frac{2 -d}{2}} (\tau')
\label{phitau}
\ee
with $\phi_\pm$ denoting $\phi(\pm \tau_0)$.
Then $\Delta \phi \equiv \phi_+-\phi_-%\phi(\tau_0)-\phi(-\tau_0)
$ is  determined as 
\be
\Delta \phi (\gamma)%\phi(\tau_0)-\phi(-\tau_0)
=\gamma\int^{\tau_0}_{-\tau_0}d\tau'  f^{\frac{2-d}{2}} (\tau') \,,
\ee
which becomes finite and a function of $\gamma$ in the end. Hence the difference $\Delta \phi \,(\in [0,\infty))$ may be used as an alternative deformation parameter replacing $\gamma$. %and, thus, the value of the scalar
As $\tau\rightarrow \pm \tau_0$ asymptotically in the far past and far future, one may easily verify 
that
\be
f= \frac{1}{(\tau_0\mp \tau)^2}+{\cal O}(1)\,, \ \ \ 
\phi=\phi_\pm %\tau_0) 
\mp %\gamma
\frac{\gamma %(\tau_0\mp \tau)^{d-1}
}{d-1}(\tau_0\mp \tau)^{d-1}+{\cal O}\bigr((\tau_0\mp \tau)^{d+1}\bigl)\,,
\label{ftau0}
\ee
which shows that the geometry is asymptotically dS in the far past and far future.
Note that there is another convenient parametrization of the time coordinate defined  by
\be
t= \int^\tau_0 d \tau' f^{\frac{1}{2}}(\tau') \,,
\label{ttau}
\ee
where the integration constant is fixed to be zero for simplicity. In this parametrization, the metric in (\ref{ansatz}) takes the form
\be \label{globalt}
ds^2= -dt^2 +f(t)\,d\Omega^2_{d-1} \,,
\ee
where $-\infty < t < \infty $.

Without the deformation ($\gamma=0$), we have the pure dS space with $\phi=\rm const.$ The equation (\ref{particle})
can be integrated to 
%\be
$f(\tau)=%\frac{1}{\cos^2 \tau}
1/\cos^2 \tau$
%\ee
with $\tau_0= \pi/2$. With $\cosh t =1/\cos \tau$, one recovers the usual form of the 
dS metric
\be \label{usual}
ds^2= -dt^2 +\cosh^2 t\, d\Omega^2_{d-1}
\ee
in  %the usual form of 
global coordinates.

% For $d=3$,  $f$ and $\phi$ 
%can be given in terms of some special functions of $\tau$ %explicitly 
%or 
%in terms of elementary functions  of the coordinate $t$ \cite{Bak:2007jm} but we shall go into any further details in that direction either.  
For $d=3$, the metric factor $f$ and $\phi$ can be given explicitly. In $\tau$ coordinate, they read \cite{Bak:2007jm} 
\bea \label{ftau}
f(\tau)&=&\frac{\kappa_+^2}{{\rm sn}^2 (\kappa_+ (\tau+\tau_0),k^2)}\,,
%=\f{\kappa_+^2{\rm dn}^2 (\kappa_+ \mu,k^2) }{{\rm cn}^2 (\kappa_+ \mu,k^2)} , %
\nonumber
\\
%\ \ \ \
\phi(\tau)
%&=& \f{1}{\sqrt{2}}\ln \left( \f{{\rm dn}(\kappa_+ (\mu+\mu_0),k^2)-k{\rm cn}
%(\kappa_+ (\mu+\mu_0),k^2)}
%{{\rm dn}(\kappa_+ (\mu+\mu_0),k^2)+k{\rm cn}
%(\kappa_+ (\mu+\mu_0),k^2)}
%\right)
%\nonumber\\
&=&\phi_0+\frac{1}{\sqrt{2}}\ln\left(\f{1+ k\, {\rm sn} (\kappa_+ \tau,k^2)}{ 1- k\, {\rm sn} (\kappa_+ \tau,k^2)}\right)\,,
%\label{phitau}
\eea
%where ${\rm sn}(x,k^2)$ is the Jacobi elliptic function.
with
\bea
%&&
\kappa^2_\pm \equiv \f{1}{2} (1\pm\sqrt{1-2\gamma^2})\,, 
%\nonumber\\
%&& 
\ \ \ k^2 \equiv \kappa^2_-/\kappa^2_+={\gamma^2\over 2}+O(\gamma^4)\,, %,\\
%&& \tau_0 = K(k^2)/\kappa_+ =\f{\pi}{2} \left(1+ \f{3}{8}\gamma^2+ O(\gamma^4)\right) .
\eea
where ${\rm sn}(x,k^2)$ is one of the Jacobi elliptic functions and $K(k^2)$ denotes the complete elliptic function of the first kind. Using the period of the Jacobi elliptic function, one finds 
\be
 \tau_0 = K(k^2)/\kappa_+ =\f{\pi}{2} \left(1+ \f{3}{8}\gamma^2+ O(\gamma^4)\right)\,.
\ee
where the expansion in $\gamma$ is consistent with (\ref{tauseries}) with $d=3$.
 In $t$ coordinate, they become
\bea\label{ftcoor}
f(t) &=& {1\over 2}\Big( 1+ \sqrt{1-2 \gamma^2} \cosh 2t \Big) \,, \nonumber\\
\phi(t) &=& \phi_0 + 
{1\over \sqrt{2}} \log\left( \frac{1+ \sqrt{1-2 \gamma^2} +\sqrt{2}\gamma \tanh t}{1+ \sqrt{1-2 \gamma^2} -\sqrt{2}\gamma \tanh t }\right) .
\eea
It is then straightforward to show that
\be \label{ephias}
%\lim_{r\to \pm \infty} 
\Delta \phi= %\pm 
{1\over \sqrt{2}} \log\left( \frac{1 +\sqrt{2}\gamma}{1- \sqrt{2}\gamma}\right)\,.
\ee

Another explicit form of metric solutions in arbitrary dimensions can be given  \cite{Bak:2016rpn} 
%in the AdS context 
but we shall not go into the details. 

%For $d=3$, $f$ and $\phi$ can be given in terms of some special functions of $\tau$ %explicitly 
%or %$f(t)$ and $\phi(t)$ can be given 
%in terms of %even simpler 
%elementary functions  
%of the coordinate $t$ 
%\cite{Bak:2007jm} but we shall go into any further details in that direction either.  

%\eq\label{fforma}
%f(r) = {1\over 2}\Big( 1+ \sqrt{1-2 \gamma^2} \cosh 2r \Big)
%\eqx
%and
%\eq
%\phi(r) = \phi_0 + 
%{1\over \sqrt{2}} \log\left( {1+ \sqrt{1-2 \gamma^2} +\sqrt{2}\gamma \tanh r  \over  1+ \sqrt{1-2 \gamma^2} -\sqrt{2}\gamma \tanh r }\right) .
%\eqx
%\eq \label{e.phias}
%\lim_{r\to \pm \infty} \phi(r) = \pm {1\over \sqrt{2}} \log\left( {1+ \sqrt{1-2 \gamma^2} +\sqrt{2}\gamma \over  1+ \sqrt{1-2 \gamma^2} - \sqrt{2}\gamma }\right) .
%\eqx

%where 
%\eqn \label{efmuone1}
%f(\mu)&=&\f{\kappa_+^2}{{\rm sn}^2 (\kappa_+ (\mu+\mu_0),k^2)}=\f{\kappa_+^2{\rm dn}^2 (\kappa_+ \mu,k^2) }{{\rm cn}^2 (\kappa_+ \mu,k^2)} , %\nonumber
%\\
%\phi(\mu)&=& \f{1}{\sqrt{2}}\ln \left( \f{{\rm dn}(\kappa_+ (\mu+\mu_0),k^2)-k{\rm cn}
%(\kappa_+ (\mu+\mu_0),k^2)}
%{{\rm dn}(\kappa_+ (\mu+\mu_0),k^2)+k{\rm cn}
%(\kappa_+ (\mu+\mu_0),k^2)}
%\right)
%\nonumber\\
%&=&\f{1}{\sqrt{2}}\ln\left(\f{1+ k\, {\rm sn} (\kappa_+ \mu,k^2)}{ 1- k\, {\rm sn} (\kappa_+ \mu,k^2)}\right)
%\label{efmuone2}
%\eqnx
%with
%\eqn
%&&\kappa^2_\pm \equiv \f{1}{2} (1\pm\sqrt{1-2\gamma^2}) ,\\
%&& k^2 \equiv \kappa^2_-/\kappa^2_+={\gamma^2\over 2}+O(\gamma^4) ,\\
%&& \mu_0 \equiv K(k^2)/\kappa_+ =\f{\pi}{2} \left(1+ \f{3}{8}\gamma^2+ O(\gamma^4)\right) .
%\eqnx
%This describes Janus deformation of the Poincare patch geometry.

\section{Observer's patches and horizons
}\label{sec3}

\begin{figure}[btp]  
\vskip0.3cm 
\begin{center}
\begin{tikzpicture}[scale=1.1]
\draw[thick,blue %,-<
](-2,0) node[left]{\color{black}$-\frac{\pi}{2}$}--++(0,2);
\draw[thick,blue] (-2,2) node[right]{$\ P$}
--++(0,2) node[left]{\color{black}$\,\,\frac{\pi}{2}$};
\draw[thick,red %, ->
](2,0)--++(0,2);\draw[thick,red] (2,2) node[left]{%\large
$P'\ $}
--++(0,2);
%\draw[blue,decoration={coil,segment length=0.5mm,amplitude=0.15mm},decorate] (-2,2) arc (180:270:2);\draw[red,decoration={coil,segment length=0.5mm,amplitude=0.15mm},decorate] (2,2) arc (0:-90:2); \draw[fill=black] (0,0) circle (0.05cm);
%\draw[decoration={zigzag},decorate](-2,0)--(2,0); 
% \draw[thick,dotted](-2,2)--++(4,0);
%\draw[decoration={zigzag},decorate](-2,4)--++(4,0); 
%%\draw[%violet,
%%decoration={zigzag,segment length=1mm,amplitude=0.5mm},decorate
%%](-2,4)--++(0.2,+0.35); 
%%\draw[decoration={zigzag,segment length=1mm,amplitude=0.5mm},decorate
%%](2.15,4)--++(-0.2,+0.35); 
%%\draw[decoration={zigzag,segment length=1mm,amplitude=0.5mm},decorate
%%](-2,0)--++(0.2,-0.35); 
%%\draw[decoration={zigzag,segment length=1mm,amplitude=0.5mm},decorate
%%](2.15,0)--++(-0.2,-0.35); 
%\draw[thick,brown,dotted] (-2,0)--++(4,4);
%\draw[thick,brown,dotted] (+2,0)--++(-4,4);
\draw (-2,0)--++(4,4);%--++(-2,2);
%\draw[dotted] (0,2)--++(2.08,2.08);\draw[dotted] (0,2)--++(2.10,-2.10);
%\draw[dotted] (0.15,2)--++(-2.10,2.10);\draw[dotted] (0.15,2)--++(-2.08,-2.08);
\draw (+2,0)--++(-4,4);%--++(2,2);
\draw (-2,0)node[below]{%\large
$\phantom{\,}
0$}--++(4,0) node[below]{%\large
$\phantom{\,}\pi$};%--++(-2,2);
\draw (-2,4)--++(4,0);
%\draw (+2,0)--++(-2,2)--++(2,2);
%\draw[blue,decoration={coil,segment length=0.5mm,amplitude=0.15mm},decorate] (6,4) arc (90:270:2);%\draw[red,decoration={coil,segment length=0.5mm,amplitude=0.15mm},decorate] (6,4) arc (90:-90:2); %\draw[fill=black] (6,0) circle (0.05cm);
 %\draw[thick,dotted](4,2)--++(4,0);\draw[fill=black] (6,0) circle (0.05cm);
%\draw[decoration={zigzag},decorate](-2,4)--++(4,0); 
%\draw[blue,decoration={coil,segment length=0.5mm,amplitude=0.2mm},decorate] (4,2) arc (180:90:2);
%\draw[red,decoration={coil,segment length=0.5mm,amplitude=0.2mm},decorate] (8,2) arc (0:90:2); 
%\draw[fill=black] (6,4) circle (0.05cm);
%\tikz[every to/.style={bend right}]
% \draw[every to/.style={bend right}] (-2,0) to (-2,4);
%%\draw[thick] (-2,0) .. controls (-1.75,2) .. (-2,4);
% \draw[every to/.style={bend left}] (2,0) to (2,4);
%%\draw[thick] (2.15,0) .. controls (1.9,2) .. (2.15,4);
\end{tikzpicture}
\end{center}
\vskip-0.3cm 
\caption{\small 
Penrose diagram for pure dS space where $P$ denotes the observer's patch at $\theta=0$ and $P'$  its complementary patch.%The left and the right cutoff trajectories are illustrated 
%as curves near the boundaries in the Penrose diagram of a typical 
%deformed space.
}\label{pureds}
\end{figure}
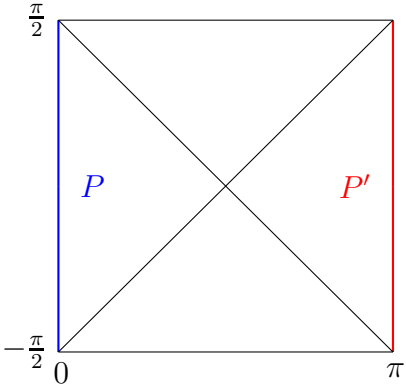 
% %
%

In order to see how the spacetime is deformed, it is illuminating to draw the Penrose diagram.
Figure \ref{pureds} shows the Penrose diagram for the pure dS space obtained with $\gamma=0$.
The horizontal and the vertical directions represent $\theta$ and $\tau$ in
\eqref{spherem} and \eqref{ansatz}, respectively, and ranged over 
$0\le \theta \le \pi$ and $-\frac\pi2 \le \tau \le \frac\pi2$. 
Each point in the diagram represents a unit $(d-2)$-sphere of the remaining directions.
We locate an observer at $\theta=0$ along the geodesic which is called the pode following \cite{Susskind:2021omt}. It is represented as the blue line in Figure \ref{pureds}.
Also, the geodesic at $\theta=\pi$ is called the antipode where another observer may be located. 
It is shown as the red line in the figure.
The future/past horizon of the pode observer is $\theta = \frac\pi2 \mp \tau$
and the static patch is defined by the interior of the horizons. It corresponds
to the left triangle denoted by $P$ in Figure \ref{pureds}. 
%Similarly,
%that of the anti-pode observer denoted by $P'$ is given by the interior of antipode's horizons.
Thus the static patches of the pode and the antipode are mutually disjoint.

\begin{figure}[tbp]   
\begin{center}
\vskip0.1cm
\begin{tikzpicture}[scale=1.1]

\coordinate (A) at (-2,2.8) ;
\coordinate (B) at (-1.4,3.4);
\fill[fill=gray!60] (A)node[left]{\color{black}$\tau$\,}--(B)--(-2,4);
\draw (-2,2.8)--(-1.4,3.4);
\draw[gray!90!black!80] (-1.4,3.4)--(-0.8,4);
\draw[thick,blue %,-<
](-2,0) node[left]{\color{black}$-\tau_0$}--++(0,2.2);
\draw[thick,blue] (-2,2) node[right]{$\ P$}
--++(0,2) node[left]{\color{black}$\tau_0$};
\draw[thick,red %, ->
](1.5,0.5) node[right]{\color{black}$\tau_0\ng-\ng\pi$}--++(0,1.5);\draw[thick,red] (1.5,2) node[left]{%\large
$P'\ $}
--++(0,1.5) node[right]{\color{black}$\pi\ng -\ng\tau_0$};
\draw (1.5,3.5)--++(0,0.5);
\draw (1.5,0)--++(0,0.5);
\draw (-2,0)--++(3.5,3.5);
\draw (+1.5,0.5)--++(-3.5,3.5);
\draw (-2,0) node[below]{$\phantom{\,}0$}--++(3.5,0) node[below]{$\phantom{\,}\pi$};%--++(-2,2);
\draw (-2,4)--++(3.5,0);
\draw[thick,blue %,-<
](5,0) node[left]{\color{black}$-\tau_0$}--++(0,2.2);
\draw[thick,blue] (5,2.2) node[right]{$\ P$}
--++(0,2.2) node[left]{\color{black}$\tau_0$};
%\draw[thick,red %, ->
%](6,2)--++(0,0.2);\draw[thick,red] (6,2.2) 
%node[left]{%\large $P' $}
%--++(0,0.2);
%\draw[thick,blue] (7,2) %node[right]{$\ P$}
%--++(0,0.4);
\draw (7,0)--++(0,2) node[right]{\color{black}$\pi\ng -\ng\tau_0$}--++(0,0.4) node[right]{\color{black}$\tau_0\ng-\ng\pi$}--++(0,2);
%\draw (7,2.4)--++(0,2);
%\draw (1.5,0)--++(0,0.5);
\draw (5,0)--++(2,2);
\draw (5,4.4)--++(2,-2);
\draw (5,0)  node[below]{$\phantom{\,}0$}--++(2,0)  node[below]{$\phantom{\,}\pi$};%--++(-2,2);
\draw (5,4.4)--++(2,0);
%\draw[fill=black] (6,4) circle (0.05cm);
\end{tikzpicture}
\vskip-0.1cm
\caption{\small 
On the left panel, we depict the Penrose diagram for the case with $\pi/2 <\tau_0 < \pi$ where $P$ denotes the observer's patch at $\theta=0$ and $P'$  its complementary patch. The shaded region describes the timelike envelope of an observer from $\tau$ to $\tau_0$. On the right, we depict the case with $\tau_0 > \pi$ where
the observer at  $\theta=0$ can see the entire Cauchy slice.
}
\label{defor}
\end{center}
\end{figure}
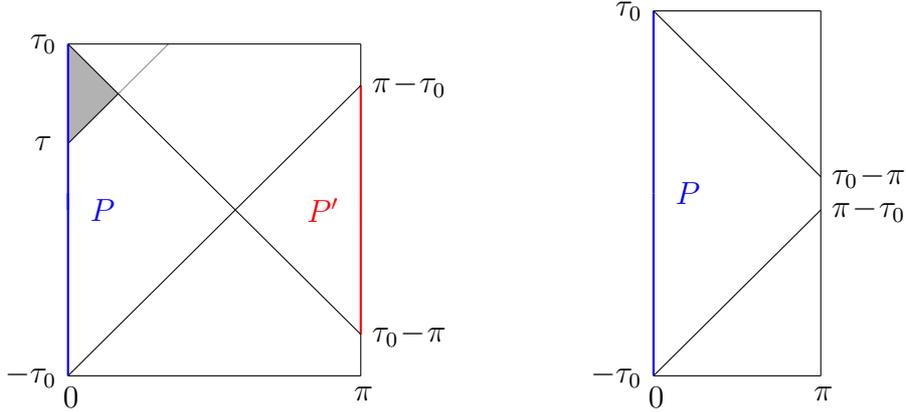

When we turn on the deformation ($\gamma \neq 0$), the range of $\tau$ is
changed to $-\tau_0 < \tau < \tau_0$ with $\tau_0> \pi/2$ where $\tau_0$ is 
a monotonically increasing function of $\gamma$, which is fully consistent with the Gao-Wald theorem~\cite{Gao:2000ga}. %, since our scalar field is an ordinary matter satisfying the null energy condition. 
It diverges to infinity
as $\gamma$ approaches the critical value $\gamma_c$ defined in 
Section \ref{sec2}. Note that the range of $\theta$ is unchanged. 
Therefore the Penrose diagram is elongated along the $\tau$ direction where the 
elongation can become infinitely large. The future/past horizon of the pode 
observer is now given by $\theta = \tau_0 \mp \tau$. The right ends of the
horizons are located at $\tau = \pm ( \tau_0 - \pi)$. 
We then have two cases depending on the value of $\tau_0$. 
If $\pi/2 < \tau_0 < \pi$, the right end of the past horizon is later than that of the future horizon as shown in the left panel of Figure \ref{defor}. 
On the other hand, if $\tau_0 > \pi$, the two horizons do not cross each other 
throughout the diagram as in the right panel of the Figure \ref{defor}.
Thus, in this case, the observer can access the entire spatial section during
the time interval $\pi - \tau_0 < \tau < \tau_0 - \pi$.

The horizon area $\mathcal{A}(\tau)$ at time $\tau$ is given by
\begin{equation}
\mathcal{A}(\tau) = \Sigma_{d-2} [\sin(\tau_0-\tau) f^{1/2}(\tau) ]^{d-2},
\label{calA}
\end{equation}
where $\Sigma_{d-2}$ is the volume of the unit $(d-2)$-sphere. 
Without the deformation ($\gamma=0$), it is a constant
$ \mathcal{A}(\tau) = \Sigma_{d-2} $. If $\gamma \neq 0$, however, 
it is monotonically increasing as a function of $\tau$ for 
$\tau_0 -\pi \le \tau \le \tau_0$, as shown in Appendix \ref{AppB}.
%which can be explicitly verified by 
%taking the derivative of $\mathcal{A}$ with respect to $\tau$.
In particular, $\mathcal{A}(\tau)$ vanishes at $\tau = \tau_0 - \pi$ 
since $f(\tau)$ has a finite value at $\tau=\tau_0 -\pi$, while
it recovers the dS value $\Sigma_{d-2}$ as $\tau$ approaches $\tau_0$
because $f(\tau)$ diverges as in \eqref{ftau0}. Thus the horizon area 
$\mathcal{A}(\tau)$ monotonically increases from 
$\mathcal{A}(\tau_0 - \pi) = 0$ to $\mathcal{A}(\tau_0) = \Sigma_{d-2}$. 
In particular, at the bifurcation horizon where $\tau = 0$, the area can
be calculated from \eqref{tauseries} and \eqref{f0series},
\begin{equation}
\mathcal{A}(0) = \Sigma_{d-2}\left[1 - \frac{\gamma^2}{2(d-1)} \right]
                + \mathcal{O}(\gamma^4)\,.
\end{equation}
The entropy of the patch $P$ is then given by $S(0) = \mathcal{A}(0)/4G$. 
At time $\tau$, the entropy of the timelike envelope represented
as the shaded area in Figure \ref{defor} is calculated as 
$S((\tau_0 + \tau)/2) = \mathcal{A}((\tau_0 + \tau)/2)/4G$.
Thus the entropy monotonocally increases in time $\tau$.
This is a manifestation of the area theorem and the generalized second law.

Now consider a scalar matter field $\chi$ of mass $m$ 
in the background metric \eqref{globalt}. It satisfies the equation of motion,
\begin{equation}
\nabla^2 \chi -m^2 \chi = 0\,.
\end{equation}
With the explicit form of the metric, it reduces to
\begin{equation}
\frac1{f^{(d-1)/2}}\partial_t [f^{(d-1)/2} \partial_t \chi]
 + \frac1f \tilde\nabla_{d-1}^2 \chi - m^2 \chi = 0\,,
\label{fnabla}
\end{equation}
where $\tilde\nabla_{d-1}^2$ is the Laplacian on a $(d-1)$-dimensional 
unit sphere. We would like to study the asymptotic behavior of $\chi$ as
$t \rightarrow \pm \infty$. From \eqref{ftau0} and \eqref{ttau}, we see that
$f$ behaves asymptotically as
\begin{equation}
f = C e^{2|t|} + \mathcal{O}(1)\,,
\end{equation}
where $C$ is some constant. In this limit, \eqref{fnabla} becomes
\begin{equation}
e^{-(d-1)|t|}\partial_t [e^{(d-1)|t|} \partial_t \chi]
 +\frac1C e^{-2|t|} \tilde\nabla_{d-1}^2 \chi - m^2 \chi = 0\,.
\end{equation}
The second term does not contribute to the leading behavior of $\chi$ 
due to the exponentially decaying factor $e^{-2|t|}$. Assuming the form
$\chi = e^{\alpha |t|}$, we find 
\begin{equation}
\alpha = -\frac{d-1}2 \pm \frac12 \sqrt{(d-1)^2 - 4m^2}.
\end{equation}
Thus, if $m\neq 0$, $\chi$ decays exponentially as $|t|$ becomes large.
In the massless case, $\chi \sim \text{const}$ is also possible in the
$|t| \rightarrow \infty$ limit. However, if one is within the sector 
where $\chi$ decays exponentially in $t \rightarrow -\infty$ limit, 
so does it in $t \rightarrow \infty$ limit as well. 

If we identify $\chi$ as a small perturbation of the massless scalar field
$\phi$ producing the Janus deformation,
\begin{equation}  \label{Pert}
\phi(x) = \phi_{cl}(t) + \chi(x),
\end{equation}
where $\phi_{cl}(t)$ is given in \eqref{phitau} with $\tau$ replaced
by $t$ via \eqref{ttau}. At least classically, we may work in the sector where we omit the perturbation that changes the initial value of the scalar field $\phi \rightarrow \phi_-$
in $t \rightarrow -\infty$ limit; thus we may work 
in the sector where
$\chi \rightarrow 0$ as $t \rightarrow -\infty$. Then in this sector we do not need to 
introduce the final state boundary condition independently, since 
it is automatically satisfied. In full quantum gravity, however, such an artificial restriction will not be allowed in general.

%Let $Y_n$ be an eigenmode of $\tilde\nabla_{d-1}^2$
%with an eigenvalue $-\lambda_n^{(d)}$, i.e.,
%\begin{equation}
%\tilde\nabla_{d-1}^2 Y_n = -\lambda_n^{(d)} Y_n.
%\end{equation}
%For example, $\lambda_n^{(3)}=n(n+1)$ and $\lambda_n^{(4)} = n(n+2)$ with
%$n=0,1,2,\ldots$. 
%
%
%Then, with $\chi = \chi_n(t) Y_n$, 
%\eqref{fnabla} becomes, in the limit $t \rightarrow \pm\infty$,
%\begin{equation}
%e^{-(d-1)|t|}\partial_t [e^{(d-1)|t|} \partial_t \chi_n(t)]
% -\frac1c e^{-2|t|| \lambda_n^{(d)} \chi_n(t) = 0.
%\end{equation}
%In this expression, we have used the asymptotic behavior of $f$, which
%can be obtained from \eqref{ftau0} and \eqref{ttau},
%\begin{equation}
%f = c e^{2|t|} + \mathcal{O}(1),
%\end{equation}

\section{Algebras and transitions
}\label{sec4}

In this section, we shall first review the algebraic approaches to dS space 
developed
in the series of papers  \cite{Witten:2018zxz,Witten:2021jzq,Chandrasekaran:2022cip,Chandrasekaran:2022eqq} and then explore related  algebraic  aspects of our Janus deformed geometry along essentially %basically 
the same direction.
% the algebraic approach to quantum field theory developed recently in de Sitter space initiated by Witten et al~\cite{Witten:2018zxz,Witten:2021jzq,Chandrasekaran:2022cip,Chandrasekaran:2022eqq}, and then we explore some aspects of  Januslike deformed geometry along this line.  

Based on some concrete examples and compelling arguments, it is now widely accepted 
that the algebra of linearized gravity plus matter 
within an open region of a  background 
spacetime  (in the limit of $G \rightarrow 0$) can be
described by
a  type ${\rm III}_{1}$ von Neumann factor
\cite{Haag:1996hvx}. 
% Recent developments in the above-mentioned papers showed
%that the incorporation of  gravity turns  the type  ${\rm III}_{1}$ 
%into a type ${\rm II}$ algebra, which can be viewed as including (perturbative) quantum gravity effects in the algebraic context.   
%The above-mentioned papers 
It is shown
that, with quantum gravity effects properly included in the algebraic context,   
the type  ${\rm III}_{1}$ algebra turns into a type ${\rm II}$ algebra \cite{Witten:2018zxz,Witten:2021jzq,Chandrasekaran:2022cip,Chandrasekaran:2022eqq}. 
Basically this structural change  
stems from the 
diffeomorphism invariance, 
which imposes rather 
strong constraints on 
admissible algebras. 
Specifically, with the static patch of pure dS space (denoted by $P$ 
in Figure \ref{pureds}), imposing the diffeomorphism  constraints makes  the algebra completely trivial. This triviality derives %is derived 
from the so-called  ``ergodic property'' of the modular automorphism group, where the 
modular Hamiltonian (operator) is given by the boost generator of the static patch.
To avoid such triviality, one may introduce 
an observer  equipped with a ``clock'' Hamiltonian\footnote{The presence of an observer in a closed universe cannot be disregarded  since the energy-momentum flux from the observer cannot  escape to infinity~\cite{Chandrasekaran:2022cip}. 
Furthermore, since a local region in spacetime may  fluctuate gravitationally, 
only the algebra accessible to a given observer can be a proper object to consider \cite{Witten:2023qsv,Witten:2023xze}.}  to relax the constraints. 
This is also natural since the static patch 
is defined with respect to the observer's worldline (in our case, the pode at $\theta=0$). In other words, the cosmological horizon and the corresponding 
patch depend on the observer.  Thus its 
 influence 
cannot be ignored even at the level of algebras. In summary,  the observer (equipped with a ``clock'')  essentially  plays a
role of  a  ``gravitational dressing'' from the perspective of the gravitational algebra
\cite{Witten:2023qsv,Witten:2023xze}. 

\begin{figure}[tbp]   
\begin{center}
\vskip0.1cm \hskip-1.5cm
\begin{tikzpicture}[scale=1.1]

\coordinate (A) at (-2,2.8) ;
\coordinate (B) at (-1.4,3.4);
%\fill[fill=gray!60] (A)--(B)--(-2,4);
%\draw (-2,2.8)--(-1.4,3.4);
%\draw[gray!90!black!80] (-1.4,3.4)--(-0.8,4);
%\fill[left,above] (-2,3.4)  node[left=1pt] {${\bf B}$};

\draw[thick,blue %,-<
](-2,0) --++(0,2.2);
\draw[thick,blue] (-2,2) --++(0,2) ;
\draw[thick,red %, ->
](1.5,0.5) --++(0,1.5);
\draw[thick,red] (1.5,2) --++(0,1.5) ;
\draw (1.5,3.5)--++(0,0.5);
\draw (1.5,0)--++(0,0.5);
\draw (-2,0)--++(3.5,3.5);
\draw (+1.5,0.5)--++(-3.5,3.5);
\draw (-2,0) -++(3.5,0) ;%--++(-2,2);
\draw (-2,4)--++(3.5,0);

\draw[purple,thick] (+1.5,2.0) node[above=-1.8pt]{\hskip-1cm\small $\color{black} \Sigma_{P'}$} --(-2.0,2.0) node[above=-1.8pt]{\hskip1.5cm\small $\color{black} \Sigma_{P}$} ;

%\coordinate (A) at (1.5,1.0) ;
\coordinate (C) at (0.0,2.0) ;
%\coordinate (B) at (-2.0,3.0) ;

\draw[purple!50] (1.5,1.0)  to [out=180, in = -37]   (C) to  [out=142, in = 0]    (-2.0,3.0) ;

\draw[purple!65] (1.5,1.5)  to [out=180, in = -36]  (C) to  [out=150, in = 0]    (-2.0,2.5) ;

\draw[purple!80] (1.5,2.5)  to [out=180, in = 36]  (C) to  [out=220, in = 0]    (-2.0,1.5) ;

\draw[purple!95] (1.5,3.0)  to [out=180, in = 37]  (C) to  [out=220, in = 0]    (-2.0,1.0) ;
\filldraw[black] (0,2) circle (0.8pt) node[below=2pt] {\tiny $\partial \Sigma$};
%\filldraw[black] (B) circle (0.8pt) node[below=2pt] {\tiny $\partial \Sigma_{\bf B}$};

\end{tikzpicture}
\vskip-0.1cm
\caption{\small 
The thick purple line ($\Sigma_{t=0}=\Sigma_{P}\cup \partial \Sigma \cup \Sigma_{P'}$ ) denotes a Cauchy surface at $t=0$ and other purple curves describe Cauchy surfaces at $t\neq 0$. $\partial \Sigma$ is for the entangling surface between the left and the right patch. 
%The boldface ${\bf B}$ denotes the time band $[\tau_{1}, \tau_{0}]$ and the   grey triangle wedge, ${\cal E}({\bf B})$, represents a timelike envelope of the time band ${\bf B}$, while $\Sigma_{\bf B}$ does the entangling surface for the wedge ${\cal E}({\bf B})$.
}
\label{deforBand}
\end{center}

\end{figure}
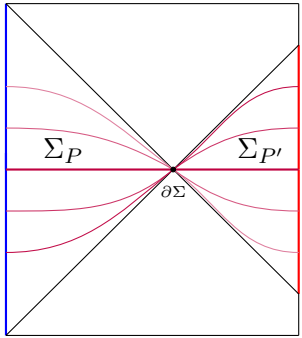

%A bit 
%Concretely, 
In dS  space, one may explicitly construct a Hilbert space ${\cal H}$ built upon the ``vacuum'' state $|\Psi_{dS}\rangle$ (known as Bunch-Davies state) that can be uniquely defined as  a  state fully invariant  under the dS isometries. 
However, shielded by the cosmological horizon,   the static-patch observer\footnote{To this observer $|\Psi_{dS}\rangle$ appears as a KMS state.} 
does not have access to the entire 
spacetime and thus to the full  Hilbert space ${\cal H}$.
%which is mainly due to the presence of the cosmological horizon.  
Thus one is restricting   the algebra to the static patch, but %However 
naively imposing the Hamiltonian gauge constraint, one gets a completely trivial algebra  as mentioned previously \cite{Chandrasekaran:2022cip}. 
%({\it i.e.} the modular Hamiltonian) as a gauge constraint, one gets a completely trivial algebra\cite{Chandrasekaran:2022cip} as mentioned above. 
 Further introducing  the observer's Hamiltonian as $H_{obs}=q$, the total Hamiltonian  within the static patch $P$ becomes $\hat{H} \equiv H_{b}+H_{obs}$, where  the bulk  Hamiltonian $H_{b}$ is  the modular Hamiltonian corresponding to  the  conserved charge associated with the boost Killing 
symmetry of the static patch, satisfying  $H_b\,|\Psi_{dS}\rangle=0$; this %Hamiltonian  
also works as  a generator for a one-parameter modular automorphism group in the algebraic context.\ng\footnote{The related %associated 
modular operator $\Delta_{\Psi_{dS}}$ is defined with respect to the vacuum state $|\Psi_{dS}\rangle$.}
 Requiring the invariance under the total Hamiltonian,  one may show that the algebra ${\cal A}_{P}$ %within the static patch $P$ 
becomes \cite{Chandrasekaran:2022cip}
%which is mainly due to the presence of the cosmological horizon.  
%Restricting then  the algebra to the static patch and imposing the boost Hamiltonian ({\it i.e.} the modular Hamiltonian) as a gauge constraint, one gets a completely trivial algebra\cite{Chandrasekaran:2022cip} as mentioned above. Introducing the observer's Hamiltonian simply as $H_{obs}=q$, the total Hamiltonian  within the static patch $P$ becomes $\hat{H} \equiv H_{b}+H_{obs}$. Requiring the invariance under the total Hamiltonian,  one may show that the bulk algebra ${\cal A}_{P}$ within the static patch $P$ becomes \cite{Chandrasekaran:2022cip}
%By restricting then the algebra on the static patch and imposing the boost Hamiltonian ({\it i.e.} the modular Hamiltonian) as a gauge constraint, one gets only a trivial algebra \cite{Chandrasekaran:2022cip} as was mentioned previously. By introducing an observer's Hamiltonian $H_{obs}=q$, and then the total Hamiltonian $\hat{H} \equiv H_{b}+H_{obs}$, it may be shown that the bulk algebra ${\cal A}_{P}$ in the static patch $P$, which is invariant under the total Hamiltonian $\hat{H}$, becomes \cite{Chandrasekaran:2022cip}
%
\begin{equation} \label{}
( {\cal A}_{0, P} \otimes B({\cal H}_{obs}) )^{\hat{H}} = \{ e^{ipH_{b}}\, a\, e^{-ipH_{b}}, ~ e^{-iqt} \}\,, \qquad [q,p]=i\,,
\end{equation}
where $p$ denotes the variable  conjugated to $q$ and  $a \in {\cal A}_{0, P}$ for any bulk  field operator within the static patch $P$.  %At this stage,
Note that 
the algebra ${\cal A}_{0, P}$ %(of the patch $P$)
acting on the Hilbert space $\cal H$ is of type ${\rm III}_{1}$ %von Neumann algebra 
as mentioned previously. 
%Here $H_{b}$ is the modular Hamiltonian, the  conserved charge associated with the boost  
%symmetry of the static patch  satisfying  $H_b\,|\Psi_{dS}\rangle=0$; it also works as a generator for a one-parameter modular automorphism group in the algebraic context\footnote{The associated modular operator $\Delta_{\Psi_{dS}}$ is defined with respect to the vacuum state $|\Psi_{dS}\rangle$.}\ng.  
This algebra ${\cal A}_{P}$ takes the standard form of  the type ${\rm II}_{\infty}$ von Neumann factor  that is obtained by the crossed product of the type ${\rm III}_{1}$ %von Neumann 
algebra ${\cal A}_{0, P}$ and the outer automorphism generated  by $H_{b}$.  To ensure  that  the observer's Hamiltonian $H_{obs}$ is bounded from below  ($H_{obs} \ge 0$), 
%it is necessary to  
one needs to project the algebra ${\cal A}_{P}$ by %using 
the projection operator $\Pi=\Theta(q>0)$. % this leads to a new algebra $\hat{\cal A}_{P} = \Pi {\cal A}_{P} \Pi$, which is %finally 
The resulting algebra $\hat{\cal A}_{P} = \Pi {\cal A}_{P} \Pi$ is
of von Neumann type ${\rm II}_{1}$ after all. %in the end.  %von Neumann algebra. 
By including the thermal energy distribution of the observer's Hamiltonian, one gets a special state  $|\Psi_{max}\rangle  \equiv |\Psi_{dS}\rangle \otimes e^{-\beta_{dS} q/2}\sqrt{\beta_{dS}}$, which may be viewed as a purification of the maximal entropy state~\cite{Chandrasekaran:2022cip}.  With this state, one can define  a  trace %in the algebra $\hat{\cal A}_{P}$ 
as 
$$
\text{Tr}_{\hat{\cal A}_{P}} \big[ \hat{a} \big]=  \langle \Psi_{max}| \hat{a} |\Psi_{max}\rangle = \int^{0}_{-\infty} \beta_{dS}\, dx~  e^{\beta_{dS} x}\,  \langle \Psi_{dS}| a | \Psi_{dS}\rangle\,, \qquad x \equiv -q
$$
where $\hat{a} =e^{ipH_{b}}\, a\, e^{-ipH_{b}}\in {\cal A}_P$.
The finiteness of the trace over $\Pi \in {\cal A}_{P}$ %or equivalently ${\bf 1}$ in $\hat{\cal A}_{P}$ 
can be easily verified, which  confirms % to classify 
that the algebra is indeed %ultimately 
of type ${\rm II}_{1}$. % with the projection. % in the end. %after all. % with the projection. % von Neumann algebra. 

On the other hand, it is argued that the algebra on any subregion 
is of  von Neumann  type ${\rm II}$ \cite{Jensen:2023yxy,Witten:2023xze,Kudler-Flam:2023qfl}.  In~\cite{Jensen:2023yxy}, one considers a subregion given by the domain of dependence ${\cal D}(\Sigma_{p})$ of the partial Cauchy surface $\Sigma_{p}$ (see for instance Figure \ref{deforBand}). The authors assume that there exists   a boost-like modular operator within the subregion, which is based on the observation that the region arbitrarily close to  
the entangling surface $\partial \Sigma_{p}$ 
 locally looks  like that of  Rindler space. 
The modular operator is further based on the existence of a KMS state $|\Psi\rangle \in {\cal H}$. Then the local property of the corresponding modular flow 
around the entangling surface $\partial \Sigma_{p}$  is used  to  construct the crossed product algebra.  Alternatively in~\cite{Witten:2023xze}, it is argued that a bulk diffeomorphism generator shifting the proper time along the observer's worldline
is sufficient to construct the invariant algebra on the timelike envelope of the observer's worldline, which is closely related to the background independence of the operator product expansion. %(as long as the observer's proper time is parametrically large compared to the background spacetime scale). 
In~\cite{Kudler-Flam:2023qfl}, the condition for the existence of a KMS state is further relaxed and instead the existence of a stationary state is assumed  
for the construction of the  modular operator. 
In these approaches,  the modular flow may not be realized as 
a geometric flow for the entire subregion 
but only a local identification of the modular flow with the corresponding geometric flow 
is sufficient to get the type ${\rm II}$ algebra.  
With the KMS (or the stationary) state $|\Psi\rangle$ as well as the thermal energy distribution of the observer's Hamiltonian, the purification of the maximal entropy state can be achieved as $|\Psi_{max}\rangle = |\Psi\rangle \otimes e^{-\beta q/2}\sqrt{\beta}$, which can be used to define a trace just as in the dS case. Unfortunately these approaches do not work for our  Janus 
deformed case  as will be shown below.

More recently, for a model with an inflaton field, it is shown that the observer's Hamiltonian 
may not be essential in constructing a non-trivial diffeomorphism invariant algebra \cite{Chen:2024rpx}.   
Of course, an observer is needed to define 
the cosmological horizon and the corresponding subregion. But the observer's Hamiltonian does not 
play any essential role 
while   the inflaton field plays the role  of  the ``clock'' in constructing nontrivial algebras. 
In~\cite{Kudler-Flam:2024psh}, the authors consider
the inflationary patch where 
the fluctuation of the inflaton field again plays the role of  the ``clock''\ng.
Interestingly, the Bunch-Davies weight~\cite{Chen:2024rpx,Witten:2023xze} can be constructed explicitly by a Euclidean method, which is not normalizable and therefore does not belong to the Hilbert space as the inflaton field becomes massless.  This situation 
is rather similar to our Janus deformed case where 
the massless scalar field is allowed  to fluctuate 
as in~\eqref{Pert}. However, in our  case, it is unclear how to construct the 
Bunch-Davies-like state or weight since the Euclidean rotation of our solution has not been fully understood. 

Although the algebraic structure appears complicated in our deformed spacetime, we expect that the Hilbert space ${\cal H}$ can be constructed, at least in principle, %on a closed universe or 
on the closed %global 
spacetime in \eqref{globalt} with a complete Cauchy surface.  The algebra in this case %is expected to 
will be described by  a von Neumann  factor of type ${\rm I}$, $B({\cal H})$, which is %consists of 
%{\it i.e.} 
the collection of bounded operators acting on the Hilbert space ${\cal H}$.   Along the  complete Cauchy surface, %$\Sigma$ 
taken as %given by 
the  time slice at $t=0$, 
%which is the time coordinate adopted in~\eqref{globalt},  
we shall assume that one can construct  
a global  state or weight $|\Omega\rangle \in {\cal H}$ just like the Bunch-Davies state or weight.
%in our Hilbert space ${\cal H}$ or at least a weight as in the case with an  inflaton field.  
The plausibility of this assumption follows from %the following 
following several considerations. First,
note that our Janus deformed geometry is smoothly connected to  dS space. %one physical motivation comes from the smooth connection of  our Janus deformed geometry  to dS space. 
It  is then natural to expect that the state (or weight) $|\Omega\rangle$ should be similar to
 the Bunch-Davies state $|\Psi_{dS}\rangle$  
%for dS space should look like  for our Janus deformed geometry
especially in the %small deformation 
limit of  
$\gamma\ll 1$.  %(see also the paragraphs around \eqref{usual}). 
Another way to see the close connection between $|\Psi_{dS}\rangle$ and $|\Omega\rangle$ is to consider the %early/late time behavior of our metric as 
$t\rightarrow \pm\infty$ limit.  Since our metric reduces to the dS metric in this limit,  it  is reasonable to 
expect  that $|\Omega\rangle$ will again look similar to $|\Psi_{dS}\rangle$ in this regime. %Second,
Furthermore, 
our Janus deformed spacetime ought to have  a state with the Reeh-Schlieder property. This is because  the existence of such a state with the Reeh-Schlieder property is proved in any globally hyperbolic 
 spacetime~\cite{Sanders:2008gs}  using the spacetime deformation argument~\cite{Fulling:1981cf,Kay:1988mu}. Of course, the same deformation argument can be applied to
 our Janus deformed case.   Finally, note that  the algebra becomes a von Neumann factor of 
type ${\rm I}$  when $\tau_{0} > \pi$, as will be argued below. Since any state can be prepared in the case of  type ${\rm I}$,  the desired  state (or weight) for $\tau_{0} \le \pi$ should  be obtainable  %near $\tau_{0} \le \pi$  
by a smooth deformation %starting 
from  the $\tau_{0} >  \pi$ counterpart. This is clearly a smooth limit, at least  from  the perspective 
of  the background geometry.

Since our deformed geometry does not allow for a boost-like Killing symmetry, we are dealing with an out-of-equilibrium state. This obscures any possible identification of the local geometric flow with a specific modular Hamiltonian and complicates the algebraic interpretation of the generalized entropy. The generalized entropy of the time envelope, represented
as the shaded area in Figure \ref{defor}, is dominated by the ${\cal O}(1/G)$ area contribution,
which increases monotonically with time reaching the dS value at future infinity as was shown in Section \ref{sec3}. This shows that the generalized entropy in our case is bounded from above. Consequently, the algebra on patch  $P$ in Figure \ref{deforBand} cannot be of von Neumann type $\rm{III}_{1}$, but should be of type $\rm{II}$, while it cannot be of type ${\rm I}_{\infty}$ with the partial Cauchy surface $\Sigma_P$. 
With the scalar fluctuation in~\eqref{Pert} after rescaling $\chi=
 \sqrt{8\pi G}\,\tilde\chi$ and keeping $\tilde\chi={\cal O}(1)$, one can follow the same argument in~\cite{Witten:2023xze} to conclude that the von Neumann algebra of the patch $P$ should be of $\rm{II}_{\infty}$. That is, one can increase the generalized entropy by turning on the background-changing mode in $\tilde\chi$. 
Since our background $\phi_{cl}$ is of order $O(1/\sqrt{G})$, the order 
 difference between $\tilde\chi$ and $\phi_{cl}$ makes the magnitude %amplitude 
of  the background-changing mode in $\tilde\chi$ unbounded as an ${\cal O}(1)$ perturbation.
It is no doubt that 
the generalized entropy can be understood algebraically  based on our state (or weight) $|\Omega\rangle$. However its explicit demonstration  is beyond the scope of the present paper.

For the case with $\tau_{0} > \pi$ (as illustrated in the right panel of Figure \ref{defor}),
the algebra becomes of type ${\rm I}_{\infty}$. This is basically due to Haag's duality\footnote{Here we assume the validity of Haag's duality in our geometry 
as ${\cal A}_{P}\vee {\cal A}_{P'} = B({\cal H})$.} as $P'= \emptyset$,  where $P'$ denotes $P$'s complementary patch in Figure~\ref{defor}. One may %It is natural to 
interpret this transition in algebra types  as  reallocation of some algebra elements from ${\cal A}_{P'} = {\cal A}'_{P}$ to ${\cal A}_{P}$ where $ {\cal A}'_{P}$ is the commutant of $ {\cal A}_{P}$.
To see what happens explicitly, let us look at the  extreme case with $\tau_{0} \gg \pi$.  
In this limit, 
the scale factor $f$ and the parameter $\gamma$ may be expanded as follows
\begin{equation} \label{}
f = f_{0} (1+ \delta f)\,, \qquad \gamma^{2} = \gamma^{2}_{c} (1-\delta_{\gamma})\,.
\end{equation}
assuming $\delta f \ll 1$.
Then the equation
\eqref{particle} reduces to 
\begin{equation} \label{}
\frac{1}{4}\Big(\frac{d}{dt}\delta f\Big)^{2} = \frac{d-1}{2} (\delta f)^{2}  - \frac{\delta_{\gamma}}{d-2} \,,
\end{equation}
which leads to the following solution 
\begin{equation} \label{}
\delta f = \textstyle{ \sqrt{\frac{2\delta_{\gamma}}{(d-1)(d-2)}}} \cosh  \sqrt{2(d-1)}\,  t\,.
\end{equation}
As a result, 
the metric takes the form
\begin{equation} \label{}
ds^{2} = -dt^{2} +\frac{d-2}{d-1}\, d\Omega^{2}_{d-1} + {\cal O}\Big(\sqrt{\delta_\gamma} e^{\sqrt{2(d-1)}|t|} \Big)\,, 
\end{equation}
%
%when %
which is vaild for the range %of $t$ 
$$
|t| \  \ll  \ \frac{1}{2\sqrt{2(d-1)}} \log \frac{2(d\ng-\ng1)(d\ng-\ng2)}{\delta_{\gamma}}\,.
$$
This shows that the spacetime for the above range of time is well approximated by a Lorentzian cylinder, which can be made arbitrarily long.  It is then clear that the corresponding spectrum of states becomes discrete and the relevant algebra is fully described by a type {\rm I} factor. % nature of the relevant algebra. 
  
%Now, some comments in order.  One can relate the difference of generalized entropy, $S_{gen} = \frac{A}{4G} + S_{bulk}$, to the relative entropy between two states $|\Omega\rangle$ and $|\Psi_{dS}\rangle$. 
%By using the relative modular operator $\Delta_{\Omega|\Psi}$, one may compute the relative entropy and relate it as 
%%
%\begin{equation} \label{}
%{\cal S}(\Omega||\Psi_{dS}) = -\langle \Omega | \ln \Delta_{\Omega|\Psi_{dS}} | \Omega\rangle  = S_{gen}(\infty) - S_{gen}(b)\,,
%\end{equation}
%%
%where $b$ denotes the entangling surface at $t=0$.
%The non-negativity of the relative entropy ensures that $S_{gen}(\infty) \ge S_{gen}(\partial\Sigma)$, whose classical version ({\it i.e.} without $S_{bulk}$)  corresponds to the area theorem. 

%of the density operator $\rho_{\Omega}$ of the state $|\Omega\rangle$  for the algebra ${\cal A}_{P}$ and the density operator  $\rho'_{\Omega}$ of the state $|\Omega\rangle$  for  ${\cal A}'_{P}$.

A comment is in order regarding the above transition of algebra types.
It will be different from that of the Hawking-Page phase transition~\cite{Hawking:1982dh}, which is algebraically again a transition %from the type
in algebra types from 
${\rm II}_{\infty}$ 
to %the type 
${\rm I}_{\infty}$ \cite{Witten:2021unn}. 
%The difference arises because, in the Hawking-Page transition,
%the  background geometry changes discontinuously, whereas the change  in our case  is smooth. 
In the Hawking-Page transition, however,
the  background geometry undergoes a discontinuous change, whereas, in our case, the change is entirely  smooth.

%%%%%%%%%%%%%%%%%%%%%%%%%%%%%%%%%%%%%%%
\section{Conclusions}\label{sec5}
%%%%%%%%%%%%%%%%%%%%%%%%%%%%%%%%%%%%%%%
In this paper we have presented a time-dependent solution to the Einstein 
equations in dS gravity coupled to a massless scalar field. It is a Janus
$\mathcal{O}(1/G)$ deformation of de Sitter space with a single deformation 
parameter associated with a time-dependent scalar field. 
It interpolates two asymptotically dS spaces in the far past and the far 
future. The Penrose diagram is elongated along the time direction, which is
consistent with Gao-Wald theorem~\cite{Gao:2000ga}. For this space, we
have explicitly shown that the entropy of the observer's patch increases
as a function of time in accordance with the area theorem.

The elongation along the time direction can be very large if we increase
deformation parameter. In this case, the neck region of the deformed space becomes a Lorentzian
cylinder. Then, we have shown that the operator algebra of the theory 
should be a type I$_\infty$ von Neumann factor. On the other hand, if
the deformation is not too large, we have argued that the algebra should
be type II$_\infty$ based on the assumption that one can construct a
global state or  weight $|\Omega \rangle$ just like the Bunch-Davies 
state/weight. We have given several plausible arguments supporting the 
existence of such a state/weight, although more rigorous treatment is needed.
Therefore, there should occur a transition in von Neumann algebra type 
from II$_\infty$ to I$_\infty$ at the point where the observer's complementary
patch disappears.

Since our deformed geometry does not have a boost %-like 
Killing symmetry,
it is not entirely clear how to identify the local geometric flow with 
a specific modular operator, which is necessary to give an algebraic 
explanation of the generalized entropy. %It would be interesting to see if it is
%possible in the present case.
This requires further investigations.

\subsection*{Acknowledgement}
We would like to thank Jong-Hyun Baek, Jeongwon Ho, and O-Kab Kwon for enlightening discussions. 
DB was supported in part by NRF Grant RS-2023-00208011, and by
Basic Science Research Program
through NRF %National Research Foundation 
funded by the Ministry of Education
(2018R1A6A1A06024977). 
%the 2023 Research Fund of the University of Seoul.
C.K.\ was supported by NRF Grant 2022R1F1A1074051.
S.-H.Y. was supported in part by  NRF-2019R1A6A1A10073079 and  by Basic Science Research Program through the National Research Foundation of Korea(NRF) funded by the Ministry of Education through the Center for Quantum Spacetime (CQUeST) of Sogang University (NRF-2020R1A6A1A03047877).
%, and  appreciates conversations and discussions with Jong-Hyun Baek, Jeongwon Ho, and O-Kab Kwon. 
%the National Research Foundation of Korea(NRF) 
%NRF Grant  %with the grant number NRF- 
%2021R1A2C1003644 and supported  by Basic Science Research Program through the NRF funded by the Ministry of Education (NRF-2020R1A6A1A03047877).

\appendix
{\center \section*{Appendix}}
%\hskip1cm

\section{Properties of the parameter $\tau_0$ %for $0\le \gamma < \gamma_c$
} \label{AppA}
%%%%%%%%%%%%%%%%%%%%%%%%%
 %%%%%%%%%%%%%%%
\renewcommand{\theequation}{\thesection.\arabic{equation}}
  \setcounter{equation}{0}
 %%%%%%%%%%%%%%%%%%%%%%%%%%
In this appendix, we shall show that the parameter $\tau_0$ begins with $\frac{\pi}{2}$ at $\gamma=0$ and
becomes monotonically increasing as a function of $\gamma$ for $\gamma \in [0,\gamma_c)$.
%and approaches infinity as one takes the limit $\gamma \rightarrow \gamma_c$.
We will also compute its leading order correction term in 
the power series expansion with respect to $\gamma$. 

We begin with %Starting from 
(\ref{tauzero}) and make a change of variable by $f= 1/g^2$.  The integral then  becomes 
\be 
\tau_0 = \int_0^{g_+} \frac{dg}{\sqrt{P(g)}}
\ee
where we introduced $g_+= 1/\sqrt{f_+}$ and
${P(g)}=1-g^2 +{\cal E} g^{2d-2}$ 
with  ${\cal E}$ denoting $\frac{\gamma^2}{(d-1)(d-2)}$.
 Using the relation $1+{\cal E} g_+^{2d-2}=g_+^2$, the above 
$P(g)$ can be rewritten as
\bea
P(g)&=& g_+^2-g^2 -{\cal E} \left( g_+^{2d-2}- g^{2d-2}\right)\nonumber\cr
&=&g_+^2 (1-q^2)\left(1- {\cal E}g_+^{2d-4} \frac{\ \,1- q^{2d-2}}{1-q^2}\right)
\eea
where $q=g/g_+$. With further change of variable from $g$ to $q$, one finds
\be 
\tau_0 = \int_0^{1} dq \frac{\ W(q)}{\sqrt{1-q^2}}%W(q)
\ee
with $W^2(q)\equiv1/\bigl(1- {\cal E}g_+^{2d-4} \frac{\ \,1- q^{2d-2}}{1-q^2}\bigr)$.
Since ${\cal E}g_+^{2d-4}$ is a monotonically increasing function of $\gamma$, $W(q)$ is monotonically increasing with $W(q) \ge 1$. Thus $\tau_0$ is monotonically increasing
as a function of $\gamma \, ( \in [0,\gamma_c)\,)$ together with $\tau_0 \ge \pi/2$. We now expand 
$W(q)$ as a power series in $\cal E$, which leads to 
\be 
\tau_0 = \frac{\pi}{2}+\frac{\cal E}{2}  \int_0^{1} \frac{dq}{\sqrt{1-q^2}}\frac{\ \,1- q^{2d-2}}{1-q^2}+ {\cal O}(\gamma^4)
\ee
where we have also used $g_+= 1+\frac{\cal E}{2} +{\cal O}(\gamma^4)$.
The above integral %in the above expression 
may be evaluated explicitly leading to the desired expression
\be
\tau_0=\frac{\pi}{2}  \left(1+ \frac{ (2d\ng -\ng 3) \Gamma(2d\ng -\ng4)}{2^{2d-4}\Gamma(d)\Gamma(d\ng-\ng 1)}\gamma^2 + {\cal O}(\gamma^4)   \right)
%\label{tauseries}
\ee 

\section{Area Theorem} \label{AppB}
%%%%%%%%%%%%%%%%%%%%%%%%%
 %%%%%%%%%%%%%%%
\renewcommand{\theequation}{\thesection.\arabic{equation}}
  \setcounter{equation}{0}
 %%%%%%%%%%%%%%%%%%%%%%%%%%
Here we shall show explicitly that the horizon area $\mathcal{A}(\tau)$
in \eqref{calA} increases monotonically for $\tau_0-\pi < \tau < \tau_0$.
With $g = f^{-1/2}$ as in Appendix \ref{AppA}, it is enough to 
consider the quantity
\begin{equation}
\sin(\tau_0 - \tau) f^{1/2} = \frac{\sin\theta}g,
\end{equation}
where we reintroduced $\theta = \tau_0 - \tau$ for convenience. Denoting 
$g' = dg / d\theta$, we have
\begin{equation}
\frac{d}{d\theta} \left( \frac{\sin\theta}g \right)
 = \frac{g \cos\theta - g' \sin\theta}{g^2} \equiv \frac{\xi}{g^2}.
\end{equation}
Therefore verifying the area theorem amounts to show $\xi(\theta) < 0$ 
for $0 < \theta < \pi$. 

Note that, from \eqref{particle}, $g$ satisfies 
\begin{equation}
g''= - g + \frac{\gamma^2}{d-2} g^{2d-3}.
\end{equation}
Then $g$ may be considered as a position of an 
anharmonic oscillator with `time' $\theta$. Since $g(\theta=0) = 0$, 
we have $\xi(0) = 0$ and
\begin{align}
\xi' &= -(g'' + g) \sin\theta \nonumber \\
     &= -\frac{\gamma^2}{d-2} g^{2d-3} \sin\theta.
\end{align}
Thus $\xi <0 $ for $0 < \theta < \pi$. This completes the proof.

\end{document}